\documentclass[useAMS,usenatbib]{mn2e}
\usepackage{graphicx}
\usepackage{color}
\usepackage{url}
\usepackage{longtable}
\usepackage [latin1]{inputenc}
\usepackage{appendix}

\definecolor{myred}{rgb}{0.7,0.0,0.2}

\definecolor{myblue}{rgb}{0.0,0.2,0.7}

\definecolor{mygreen}{rgb}{0.2,0.7,0.0}

\title[WR companions]{A Speckle-Imaging Search for Close and Very Faint Companions to the Nearest and Brightest Wolf-Rayet Stars}
\author[M. Shara et al.]{Michael M. Shara$^{1}$\thanks{E-mail: mshara@amnh.org}, Steve B. Howell$^{2}$, Elise Furlan$^{3}$, Crystal L. Gnilka$^{2}$,
\newauthor{Anthony F.J. Moffat $^{4}$,  Nicholas J. Scott$^{2}$, and David Zurek$^{1}$}
\\
$^{1}$Department of Astrophysics, American Museum of Natural History, Central Park West at 79th Street, New York, NY 10024, USA\\
$^{2}$NASA Ames Research Center, Moffett Field, CA 94035, USA\\
$^{3}$NASA Exoplanet Science Institute, Caltech/IPAC, Mail Code 100-22, 1200 E. California Blvd., Pasadena, CA 91125, USA\\
$^{4}$D\'epartement de Physique et Centre de Recherche en Astrophysique du Qu\'ebec, Universit\'e de Montr\'eal, Montr\'eal, QC H3C 3J7, Canada
}

\begin{document}

\date{Accepted  Received}


\maketitle


\begin{abstract}
Gravitationally bound companions to stars enable determinations of their masses, and offer clues to their formation, evolution and dynamical histories. So motivated, we have carried out a speckle imaging survey of eight of the nearest and brightest Wolf-Rayet (WR) stars to directly measure the frequency of their resolvable companions, and to search for much fainter companions than hitherto possible. We found one new, close companion to each of WR 113, WR 115 and WR 120 in the separation range $\sim$ 0.2" - 1.2".  Our results provide more evidence that similar-brightness, close companions to WR stars are common. More remarkably, they also demonstrate that the predicted, but much fainter and thus elusive companions to WR stars are now within reach of modern speckle cameras on 8m class telescopes by finding the first example. The new companion to WR 113 is just 1.16" distant from it, and is $\sim$ 8 magnitudes fainter than the WR star. The empirical probability of a chance line-of-sight of the faint companion at the position of WR 113 is $<$ 0.5\%, though we cannot yet prove or disprove if the two stars are gravitationally bound. If these three new detections are physical companions we suggest, based on their narrowband magnitudes, colours, reddenings and GAIA distances that the companions to WR113, WR 115 and WR 120 are an F-type dwarf, an early B-type dwarf, and 
a WNE-type WR star, respectively.
\end{abstract}



\section{Introduction and Motivation}

\subsection{Wolf-Rayet stars} 

The descendants of O stars are Wolf-Rayet (WR) stars, whose large initial masses (typically 20 - 100 M$_\odot$ at
birth), short lifetimes (typically $\sim$ 300 kyr) and strong emission lines make them excellent tracers of recent star formation
\citep{cro07}. Single Nitrogen-rich (WN type) and Carbon-rich (WC and even hotter but rare WO types) WR stars likely die
as type Ib \citep{gas86} and Ic \citep{cor12} supernovae, respectively. A large fraction of massive stars are born in binaries 
that eventually exchange mass (see references below), so that a major fraction of type Ib and Ic SNe may be produced by 
the low mass, stripped cores of formerly massive stars \citep{nom95,wos95,tau15,pre19,des20,wos21}. WR stars 
are thus important to both stellar astrophysics and supernova theory. 

The intense and inhomogeneous stellar winds of WR stars ($>$10$^{-5}$ M$_\odot$/yr \citep{bar81,shm89,hil91,nug98,hil98,vin05,pul08,smi14,san18,pon21}) 
add a significant amount of C and O, and some N, to the interstellar medium (ISM), and contribute a significant fraction of the Galaxy's ISM energy and momentum
budgets. WR stars display extremely strong emission lines (up to 1000 $\AA$  Equivalent Width),
which are probes of the rapidly expanding ($\sim$ 3000 km/s \citep{und62,abb78}) and blob-filled \citep{mof88} atmospheres 
of these very hot, evolved stars.

\subsection{OB and WR star binaries}
The single most important parameter controlling the evolution of any star is its mass. The vast majority of accurately determined stellar masses are dynamical, based on double-lined spectroscopic binaries that are either eclipsing, resolved or have orbital inclinations determined via polarimetry or colliding wind geometry. The most massive stars known, those of O-type, show the highest frequency of binarity \citep{moe19}, in excess of 50\% \citep{mas98,san12}.  At present, virtually every O star companion known is another O or early B-type star, or a WR star.  Over time the initial binary fraction will decrease because some stars explode as supernovae, or because of large scale mass loss via winds and/or nonconservative Roche lobe overflow. But a significant fraction of O-type stars' descendants, the WR stars, are still expected to be located in binaries with either O-type stars or other WR stars. It is these binary WR stars that allow the mass determinations so critical in evaluating their evolutionary histories and eventual fates.

The first WR spectroscopic binaries were detected eight decades ago \citep{wil39,wil41}, at the same time as the first close, visually resolved WR star \citep{wil40} and the first eclipsing WR binary \citep{gap41}. By 1968 a catalog \citep{smi68} of 127 Galactic WR stars had been assembled, of which 40 were certain binaries. \citet{kuh73} pointed out that an astonishing 70\% of those WR stars brighter than 10th magnitude (and presumably the closest and best studied of the WR stars) in the \citep{smi68} catalog are binary. \citet{van80} noted a smaller than 40\% incidence of OB companions to then-known Galactic and LMC WR stars, but a 100\% incidence for SMC WR stars. \citet{bar01} found a low binary frequency ($\sim$13\%) for LMC WC/WO stars.
Systematic searches for orbital radial velocity variations and the frequent presence of blue-shifted single WNh-star absorption lines, originally mistaken for O-type stars in WN+O binaries, demonstrated the binary fraction in the SMC to be similar to that in the LMC and the Galaxy: $\sim$ 40\% \citep{foe03,sch08}. 

By 2001 the Galactic WR star census had increased to 227,\footnote{Currently, the continuously updated on-line Galactic WR catalogue of Crowther (http://pacrowther.staff.shef.ac.uk/WRcat/) contains 667 stars.} with the certain and near-certain binary fraction at 39\% \citep{van01}. This is moderately lower than O-star binarity \citep{san12,cab20}, as expected if supernovae converted some massive binaries into single stars.

New tools to detect binarity in WR and OB stars (non-thermal radio emission, excess X-ray fluxes, and episodic/periodic IR excesses) were introduced in the 1980s and 1990s \citep{van93}.
Groundbased speckle imaging with 4m class telescopes, capable of detecting companions within $\sim$ 3 magnitudes of the primary, and as close as 40 mas from it \citep{har99,mas09}, detected 134 companions to 561 O, B and Wolf-Rayet stars, i.e. 24\% of the massive star sample. Interferometry carried out via the {\it Hubble Space Telescope} Fine Guidance Sensors resolved binary systems with separations as small as 10 mas and magnitude differences as large as $\Delta$V = 4 mag \citep{cab14,ald15}. These studies determined binary fractions of 22-26\% for 58 of the brightest members of Cyg OB2, and 29\% for 224 Galactic OB and luminous Blue Variable stars. Direct imaging with the Hubble Space Telescope \citep{nie98,wal01,wal12} enabled detections of companions of WR stars to as close as $\sim$ 50 mas, with brightness differences up to 3 magnitudes, and similar binary fractions. 

Amongst the most recent surveys, a JHK adaptive optics survey used the Gemini Near-infrared imager to search for close companions of OB stars in the nearby Cyg OB2 association.
Equal-brightness pairs as close as 80 mas could be resolved, and progressively fainter companions were detectable out to $\Delta$K = 9 mag at a separation of 2". Close companions to 47\% of 74 massive OB stars in the Cygnus OB2 association \citep{cab20} were found. This doubling of the same authors'  FGS-determined binary fraction is mostly due to the much larger dynamic range of the Gemini study. These Gemini results complement the radial velocity survey of \citet{kob14} (and references therein) who searched for short period, spectroscopic systems in Cyg OB2. They determined that 30\% of their sample are spectroscopic binaries with periods less than 45 days. The two combined surveys' derived multiplicity fraction is MF = 0.65 $\pm$ 0.05 and the companion frequency is CF = 1.11 $\pm $ 0.13, even without accounting for systems in the relatively unexplored separation range of 1 to 100 AU. The overall binary fraction of OB stars in associations is clearly very high.

The current state-of-the-art in bright WR binary star resolution is the CHARA interferometer, which is resolving WR companions closer than 1 mas. This has recently resulted in well-resolved orbits for two previously known binaries: WR 137 and WR 138  \citep{ric16}, and in addition, solidly determined WR and O star masses for the already-known binaries WR 140 \citep{tho21} and WR 133 \citep{ric21}. Unfortunately, CHARA is limited to stars brighter than H $\sim$ 8.5, and the vast majority of WR stars are fainter than this limit, so it cannot be the ``go-to" instrument in searches for new or faint companions.

\subsection{Motivation and outline of this study} 

Virtually all known close companions to WR stars are other WR stars, or OB stars no more than 2-3 magnitudes fainter than those WR stars \citep{cro20}. Is there some fundamental reason that prevents WR stars from existing in binaries with much fainter and lower mass companions? Or are such companions common, but undetected, both because they are so much fainter than their luminous primaries, and because they are outnumbered by faint, line of sight field stars? Because no intrinsically faint companions to WR stars are known we turn to their progenitors, the OB stars, for guidance.

\citet{moe17} compiled a wide range of OB stars' empirical binary properties. They determined that binaries with small separations (a $<$ 0.4 AU) favor modest average mass ratios q $\sim$ 0.5; binaries with intermediate separations (a $\sim$ 10 AU) have mass ratios weighted toward small values q $\sim$ 0.2 - 0.3; while widely separated companions (a $\sim$ 200 - 5,000 AU) are outer tertiary components in hierarchical triples with q throughout the range 0.1 - 1.0 .  

Numerical simulations of the accretion of gas onto forming binary stars suggested that $\sim$ equal mass binaries are preferred \citep{bat97}. This is because the secondaries in forming binaries tend to be more efficient at accreting gas from the binary accretion disk as they have wider orbits. This tends to equalize the binary components' masses, predicting that WR stars' usual companions should be massive, luminous stars... as is observed.

More recently, however, numerical simulations also demonstrate that lower mass stars with luminosities far smaller than those of WR stars can be dynamically captured by O stars \citep{wal19} during the first few million years of star cluster evolution. In addition, coupled magnetohydrodynamic and N-body simulations have recently shown \citep{cou21} that dynamical interactions between stars in the presence of gas during cluster formation modify the initial distributions towards binaries with larger mass ratios. This mechanism, which also predicts the presence of low mass companions to massive stars, is consistent with the findings of \citet{moe17}. 

Many of the $\sim$ 60\% of apparently single WR stars \citep{cro20} could have companions an order of magnitude less massive and 1000X fainter than themselves, and we would be unaware of their existence. Spectroscopic and speckle surveys to date have been insensitive to any such lower luminosity companions. Their existence, or not, is a direct test of dynamical formation of very unequal mass binary and triple stars early in the lives of star clusters.

A new generation of speckle cameras, operating at visible wavelengths and mounted on 8 meter class telescopes is now available. Their high resolution imaging capabilities can detect companions much fainter than hitherto possible, and much fainter than the CHARA limits: 5 (10) magnitudes fainter than O or WR primaries, at separations of 17 (1000) mas, respectively, corresponding to the range of large separations where OB stars sometimes display faint, distant companions \citep{moe17}. We are undertaking a systematic search for such faint companions amongst the closest and brightest known WR stars, and report our first results here, which demonstrate the power of the newest speckle cameras coupled to 8 m class telescopes.

In Section 2 we describe the instrument used for our observations. Our targets and observations are listed in section 3. In section 4 we show our observational results, including detections of previously unknown, close companions to three of the eight WR stars we observed. We briefly summarize our results in section 5.
 
\section{The cameras and filters} 
We used the speckle camera `Alopeke located at the Gemini-N 8-meter telescope on Mauna Kea, Hawaii. The detector of `Alopeke is a dual-channel imager using two electron-multiplying CCDs (EMCCDs). `Alopeke provides simultaneous two-color, diffraction-limited optical imaging (FWHM $\sim$ 0.015" at 466 nm) of targets over a 6.7" field-of-view as faint as V $\sim$ 17. Very close companions (0.02" - 0.1") with magnitude differences as large as 5 magnitudes, and wider companions (0.1"-1.2") with very large magnitude differences (up to 8-10 mags) can be resolved. The camera has filter wheels providing bandpass limited observations. 

Multiple sets of 1000 exposures of 60 msec each comprise each observation, as well as calibration observations. See \citet{sco18,sco21} for a detailed description of the instruments and their filters, as well as the `Alopeke-Zorro Web pages{\footnote{https://www.gemini.edu/instrumentation/alopeke-zorro}}.  See \citet{hor12} and \citet{how11} for descriptions of the data reductions and the final data products, respectively. 

We deliberately chose the 466 nm and 716 nm narrowband filters for the blue and red-side imagery. These filters have full widths at half maximum (FWHM) of 44 and 52 nm, respectively. These blue and red filters include, respectively, the HeII 468.6nm (CIII 465.0/CIV 465.8nm) lines in WN (WC) stars, and the CIV 706nm (NIV 712nm) lines of WC (WN) stars, in addition to the weaker 706nm line of HeI which is strongest in later type WR stars. These emission line filters should be helpful in the detection of close WR binary companions and/or ionized circumstellar helium.

\section{Targets and Observations} In this first, exploratory study carried out in June 2020, we observed eight of the closest and best-studied WR stars. All targets' distances range from 1.3 to 2.7 kpc. Three of the eight targets were already known spectrographically to have close O-type companions, but all three were expected to be much closer than the limit of resolution of `Alopeke. In Table 1 we list these eight targets, their WR subtypes, line-free b and v magnitudes, Gaia-determined distances and proper motions, the dates of observation and integration times.

\section{Results}

In Table 2 we again list all the WR stars of Table 1, and report on each detected companions' magnitude difference, position angle, and angular and projected linear distances from each WR star in each filter's observations.

The results of all observations are shown in Figures 1 through 4. The red and blue curves in each figure correspond to the two filters used in each observation. They measure the 5-$\sigma$ contrast limits $\Delta$m achieved in magnitudes fainter than the WR star. The reconstructed speckle images, covering a $\sim$ 1.5"x1.5" region centered on each WR star are also shown as insets. In a few cases the surrounding background appears asymmetric and/or ``ringlike". These ``features" are at the noise limit of the images, and not real. Where only one reconstructed image inset is shown, the blue image was too noisy for any reliable detection.

The key result of this paper is that previously undetected, close companions were found for three of the eight observed target WR stars (WR 113, WR115 and WR 120) in the separation range 0.2" - 1.2". Details of the detected companions, and limits on non-detections follow.

\subsection{WR 110}

WR110 = HD 165688 is an apparently single WN5-6b WR star which displays a 4.08 $\pm$ 0.55 day periodicity with amplitude $\sim$ 0.01 mag \citep{che11}. The small amplitude brightness variations are conjectured to arise in the inner parts of a corotating interaction region seen in emission as it rotates around with the star. There is no indication of any companion within 100 mas of WR 110 that is within $\sim$ 3 (5) magnitudes of the WR star in the 466 nm (716 nm) narrowband filter speckle images. Neither is any companion detected within 4 (7) magnitudes in the 466 nm (716 nm) filter images out to 1.2" from WR 110. These two angular separations correspond to projected linear separations of 178 and 2136 AU, respectively.

\subsection{WR 111}

WR 111 = HD 165763  is an apparently single WC5 WR star which shows no periodicities with amplitudes larger than $\sim$ 3 mmag, measured every $\sim$ 11 sec over a 3 week period \citep{mof08}.  There is no indication of any companion within 100 mas of WR 111 that is within $\sim$ 3 (4.5) magnitudes of the WR star in the 466 nm (716 nm) narrowband image filters. Neither is any companion detected within 4 (8) magnitudes in the 466 nm (716 nm) image filters out to 1.2" from WR 111. These two angular separations corresponds to projected linear separations of 131 and 1572 AU, respectively.

\subsection{WR 113}

WR 113 = HD 168206 = CV Serpentis is a well-known and intensely studied WR star binary, surrounded by an extensive ring nebula \citep{mil93,cap02}. The WC8d WR star and its O8-9IV companion orbit each other in just 29.7 days \citep{cow71,mas81,dav12}. The well-determined orbital inclination and radial velocity orbits \citep{hil18} lead to some of the best determined massive star masses known: 11.7 $\pm$ 0.9 M$_\odot$ for the WR star and 33.3 $\pm$ 2.0 M$_\odot$ for the O star. The maximum projected linear separation of the O and WR stars is $\sim$ 129 R$_\odot$ \citep{hil18}, corresponding to an angular separation of $\sim$ 680 microarcsec, far too small for  `Alopeke + Gemini-N to resolve. But we do resolve a hitherto unknown companion at 1.16" (corresponding to a projected linear distance of $\sim$2200 AU from WR 113) that is $\sim$ 8 mag fainter than the WR star. The faint v $\sim$ 18 mag companion is clearly detected, but only in the 716 nm images, which are much (3-4 magnitudes) more sensitive than the 466 nm images (see Figure 3). Because of this much greater red sensitivity the companion is not necessarily red. 

The apparent magnitude of WR 113 is G$\sim$8.8 mag so the companion must have G$\sim$16.8 mag. The distance modulus of WR 113 is $\sim$11.4 and its G-band absorption is $A_G$ $\sim$2.5 \citep{rat20}. If the WR star and newly resolved companion are physically associated, the absolute magnitude of the companion is then $\sim$+2.9. This suggests that it is of spectral type F and on the main sequence so that any bolometric correction will be small. This would be the intrinsically lowest luminosity ($\sim 5 L_\odot$) companion known to any WR star.

If the companion is physically associated with the O + WR star binary, Kepler's third law determines that its orbital period with respect to the massive binary must be of order $10^{5}$ years. In that case the separation and position angle of the companion will not change perceptibly over the coming decades. If the companion's angular separation and/or position angle do change significantly in the coming decades because of proper motion then the companion is not physically associated with WR 113. A good optical or NIR spectrum of the companion to the WR star would yield a spectrographic parallax, which could also rule out a physical association. This applies for all three newly detected companions, but will be challenging in all cases because of the small angular separation of each WR star from its newly detected companion.

\subsection{WR 115}

WR 115 = HIP 90299 is an apparently single WN6o WR star which displays variability ascribed to a corotating Interaction region \citep{stl09}, leading to structures caused by perturbations at the base of the WR star wind that propagate through it and are carried by rotation. WR 115 also emits very hard X-rays \citep{ley07}. It displays diluted emission lines (d.e.l.), which hints at binarity, but d.e.l. is by no means a definitive diagnostic \citep{ham07}. 

Our speckle observations clearly resolve a previously unknown companion star just 0.20" from WR 115, and 1.03 (1.60) mag fainter than it in the 716 (466) nm images. The companion is faint in the raw and reconstructed 466 nm images, but seen at the same angular separation and position angle as in the 716 nm images, so the detection is confirmed. The clearest detection is seen when using the comparison point source HR 8852 (a B9V star) during the processing of the speckle data, though this results in a slight (obviously nonphysical) elongation of the WR star and companion in the reconstructed image. 

An intrinsically blue (but non-WR star) physical companion to the WR star would be fainter in the blue filter than the red one owing to the excess emission from 4686 HeII in the WR star... as is observed. We observe that the companion is 1.6 mag fainter in the b filter than WR115, so its b = 15.0 mag. The WR 115 distance modulus DM = 10.65, its reddening is $E_{b-v}$ = 1.5 \citep{ham19}, its absorption $A_{v}$ = 4.1 $E_{b-v}$ = 6.1, and $A_{b}$/$A_{v}$ $\sim$1.15 , thus $A_{b}$ = 7.0, and the companion displays $M_{b}$ = -2.65. This luminosity suggests that the companion, if physical, is an early B dwarf.

This new companion is unresolved from the WR star in all ground-based spectroscopy, and would account for WR 115's diluted emission lines as well as its likely-spurious Gaia parallax with high astrometric excess noise. If physically associated with the WR star, the companion's projected linear separation is $\sim$270 AU from WR 115. If the stars are not physically associated then the relatively large Gaia-determined proper motion of WR 115 (9.1 mas/yr) will noticeably change the separation and position angle of WR 115 and its companion within a few years. 

\subsection{WR 120}

WR120 = V462 Sct is an apparently single WN7o WR star which displays variability ascribed to a corotating Interaction region \citep{stl09}. Our speckle observations clearly resolve a previously unknown companion in both filters. The nearly identical separation and position angle of that companion in both filters' images confirm its reality. The much higher S/N observations in the red filter yield an angular separation of $\sim$ 0.66". This angular separation corresponds to a projected linear separation of $\sim$ 1800 AU.

All observations were reduced with 10 different (single) comparison stars that served as point source calibrators. The 466 nm reconstruction shown in Figure 3 (using point source calibrator HR 7174) results in a somewhat elongated shape for the primary star, but the other point source calibrators observed during the same night did not yield better results. The companion is faint in the lower S/N blue image. The average (using 10 comparison stars) magnitude differences between the WR primary and the newly detected companion are 2.11 (at 466 nm) and 3.55 (at 716 nm) mag, respectively, with larger ($\pm$0.3) uncertainties for the 466 nm data. If the companion is a WR star with much stronger emission in the 466 nm filter than that of WR 120 then its blue excess is explainable. 

We observe that the companion is 2.1 mag fainter in the b filter than WR120, so its b = 15.4 mag. The WR 120 distance modulus DM = 12.2, its reddening $E_{b-v}$ = 1.25 \citep{ham19}, its absorption $A_{v}$ = 4.1 $E_{b-v}$ = 5.1, and $A_{b}$/$A_{v}$ $\sim$1.15 , so $A_{b}$ = 5.9.  The companion thus displays $M_{b}$ = -2.7. This luminosity and the strong blue excess at 466 nm suggests that the companion, if physical, is a WN3-WN4 WR star. 

The modest proper motion of WR 120, 3.25 mas/yr, will noticeably separate the Wolf-Rayet star from its companion in about a decade if they are not gravitationally bound. If the companion is line-of-sight and much closer than the WR star then a noticeably varying separation and/or position angle will occur more quickly.

\subsection{WR 134}

WR 134 = HD 191765 is an apparently single WN6b WR star that does not display diluted emission lines. 
An extensive spectroscopic campaign \citep{ald16} determined the corotating Interaction region period as 2.255 days.
There is no indication of any companion within 100 mas of WR 134 that is within $\sim$ 3 (4.5) magnitudes of the WR star in the 466 nm (716 nm) narrowband image filters. Neither is any companion detected within 4 (8) magnitudes in the 466 nm (716 nm) image filters out to 1.2" from WR 134. These two angular separations correspond to projected linear separations of 182 and 2184 AU, respectively.
\subsection{WR 139}

WR 139 = HD 193576 =V444 Cyg is a double-lined, eclipsing spectroscopic binary with WN5o and O6 III-V components in a 4.2 day orbital period \citep{mar97} with well-determined orbital parameters and stellar masses \citep{eri11}. The components' colliding winds \citep{sho88,lom15} are well-studied, and display asymmetric behavior around the eclipses in the system's X-ray light curves. The large opening angle of the X-ray emitting region is evidence of radiative braking/inhibition occurring within the system. Polarimetry \citep{rob90,stl93} shows evidence of the cavity the wind-wind collision region carves out of the Wolf-Rayet star's wind. 

There is no indication of any companion within 100 mas of WR 139 that is within $\sim$ 5 magnitudes of the WR star in either the 466 nm or 716 nm narrowband reconstructed images. Neither is any companion detected within 7 (8.6) magnitudes in the 466 nm (716 nm) images out to 1.2" from WR 139. These two angular separations correspond to projected linear separations of 134 and 1608 AU, respectively.

\subsection{WR 140}

WR 140 = HD 193793 is a double-lined spectroscopic binary with WC7pd and O5.5fc components \citep{fah11}. It is well resolved with the CHARA interferometer, which has yielded an exquisite orbit and masses for the component stars \citep{tho21}. The O and WR stars of WR 140 were below the resolution limit of `Alopeke+ Gemini-N when we observed them in June 2020. The star appears slightly elongated in the speckle images, but the binary was not resolved. There is no indication of any well-resolved companion within 20-100 mas of WR 140 that is within $\sim$ 5 magnitudes of the WR star in either the 466 nm or 716 nm narrowband image filters. Neither is any companion detected within 7 (8.6) magnitudes in the 466 nm (716 nm) image filters out to 1.2" from WR 140. These two angular separations correspond to projected linear separations of 178 and 2136 AU, respectively.

\subsection{Are these three new detections physical companions?}

We have carried out sensitive and high angular resolution speckle observations of eight of the closest WR stars. WR 115  and WR 120 have close (0.20" and 0.66") and hitherto undetected companions (1.6 and 2.1 mag fainter than themselves at 466 nm). A third companion (to WR 113, and 1.16" distant from it) is 8 magnitudes fainter than the WR star. 

Our companion detection rate (for the two closer companions ($\rho < 0.7"$) of luminosity comparable to their WR stars) is 25\%. This is in good accord with the previous speckle and HST interferometric and imaging studies described in section 1.2. No very faint companions to WR stars are currently known, so our one-in-eight detection (of the companion to WR 113) cannot be compared with the results of previous searches.

As already noted in the Introduction, numerical simulations suggest that $\sim$ equal mass binaries are preferred during binary star formation \citep{bat97}. However, dynamical captures and/or exchanges of low mass stars are also predicted \citep{wal19,cou21}, so that some low mass, low luminosity companions to O and WR stars are predicted to exist amongst present or former star-formation region or star cluster members. It has recently been shown that the fraction of WR stars currently in such regions is low (\citet{rbt20} and references therein). In particular, WR 110 and WR 111 are not members of Sgr OB1; WR 134 is not a member of Cyg OB3; and WR 139 is not a member of Cyg OB1 or Berkeley 86. 

We are unaware of any cluster or association that WR 113 or WR 115 might be a member of, and inspection of Digitized Sky Survey plates shows no clustering anywhere near them. WR 120 is a possible member of Dolidze 33. 

The stars best-studied to find much fainter companions using `Alopeke are solar-type stars \citep{hor14,mat18} and exoplanet hosts \citep{les21}. These stars are far less massive and luminous than our eight WR targets, so their dynamical histories are likely to have been very different from those of WR stars. The \citet{mat18} results suggest that a large majority of arcsecond and closer companions within $\sim$ 5 magnitudes of such solar-type stars are gravitationally bound, while modeling suggests that fainter companions are more likely line-of-sight. 

To empirically check on the likelihoods that these new companions are field stars, we queried the Gaia EDR3 database \citep{gai21} for all stars within 200" of, and centered on WR 113, WR 115 and WR 120.

Within a radius of 200" of WR 113, i.e. within the 125,664 $arcsec^{2}$ surrounding WR 113, there are 88 (144) stars brighter than g = 17 (18). The empirically determined {\it a priori} probability of locating a g $<$ 17 (18) magnitude star within 1.16" of the position of WR 113 is then just $\sim $ 0.3\% (0.5\%).  No previous speckle imaging (or any other technique) survey of massive and luminous stars exists that could detect such companions. If the companion to WR 113 is gravitationally bound, it would be the lowest luminosity ($\sim 5 L_\odot$) companion known to any WR star. A main sequence companion of that luminosity would be of F spectral spectral type and have a mass of $\sim$ 1.7 M$_\odot$.

The apparent B-band brightness of WR 115 is B = 13.09. WR 115's companion is 1.6 mag fainter than it in the 466 nm filter, so we searched for stars with G$_{BP}$ $<$ 15 (16) within a radius of 200" of WR 115; there are 18 (31) such stars. The empirically determined {\it a priori} probability of locating a G$_{BP}$ $<$ 15 (16) mag star within 0.2" of the position of WR 115 is just $\sim $ 0.2\% (0.3\%).  

The apparent B-band brightness of WR 120 is B = 12.97. WR 120's companion is 3.5 mag fainter than it in the 466 nm filter, so we searched for stars with G$_{BP}$ $<$ 17 (18) within a radius of 200" of WR 120; there are 51 (113) such stars. The empirically determined {\it a priori} probability of locating a G$_{BP}$ $<$ 15 (16) mag star within 0.66" of the position of WR 115 is just $\sim $ 0.2\% (0.3 \%). 

All of these probabilities are, however just that - {\it a posteriori} probabilities. Fortunately, a definitive determination of whether each of these companions is gravitationally bound to its nearby WR star is straightforward. A second epoch of speckle observations of the same targets on timescales of just a few years will reveal varying separations and/or position angles due to different proper motions of line-of-sight companions, or the constant separations and position angles of physical companions (whose orbital periods are $>$ 10,000 yrs). 

\clearpage
\begin{figure}

\hspace{4cm}
\includegraphics[width=10 cm]{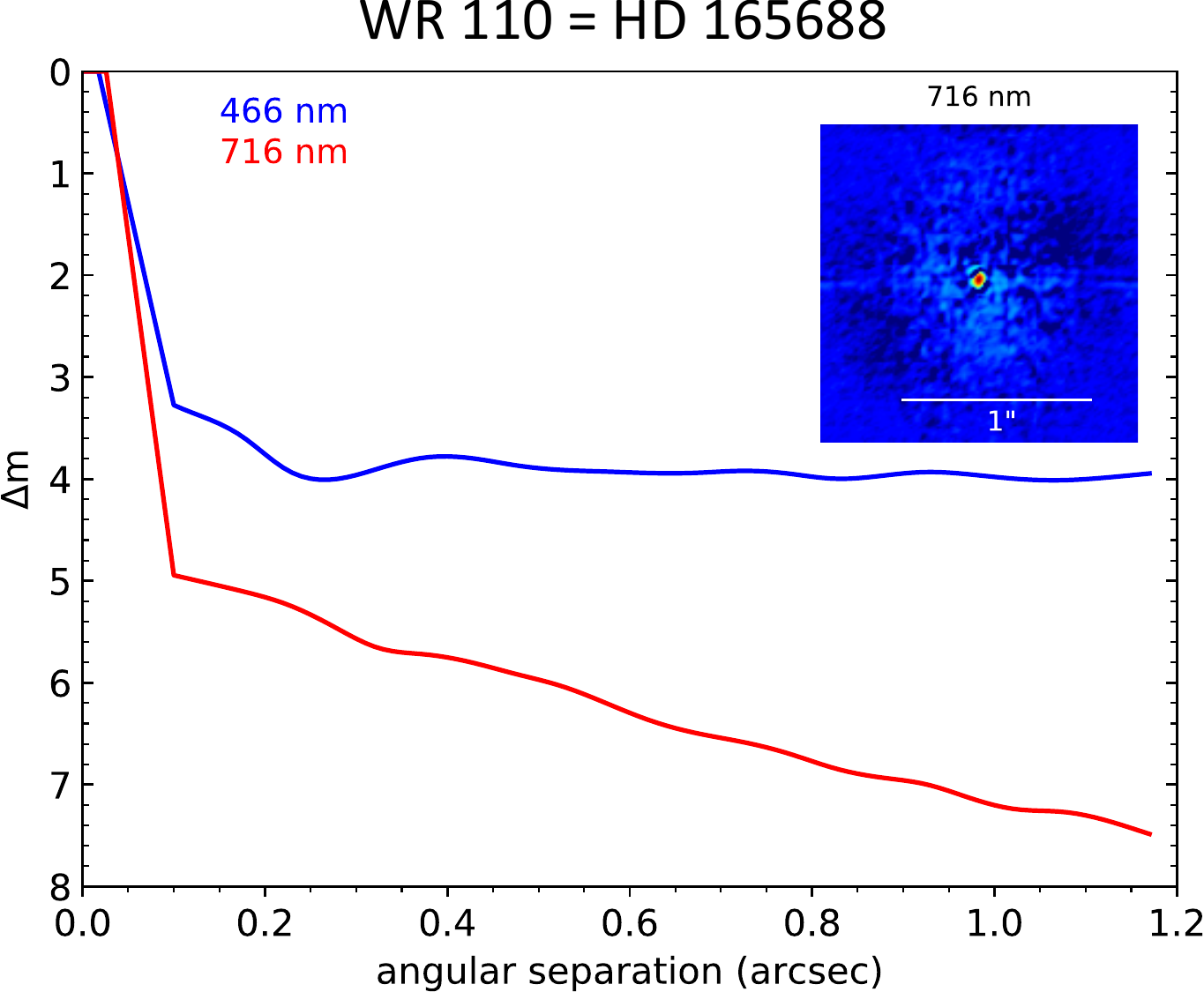}

\vspace{2mm}
\hspace{4 cm}
\includegraphics[width=10 cm]{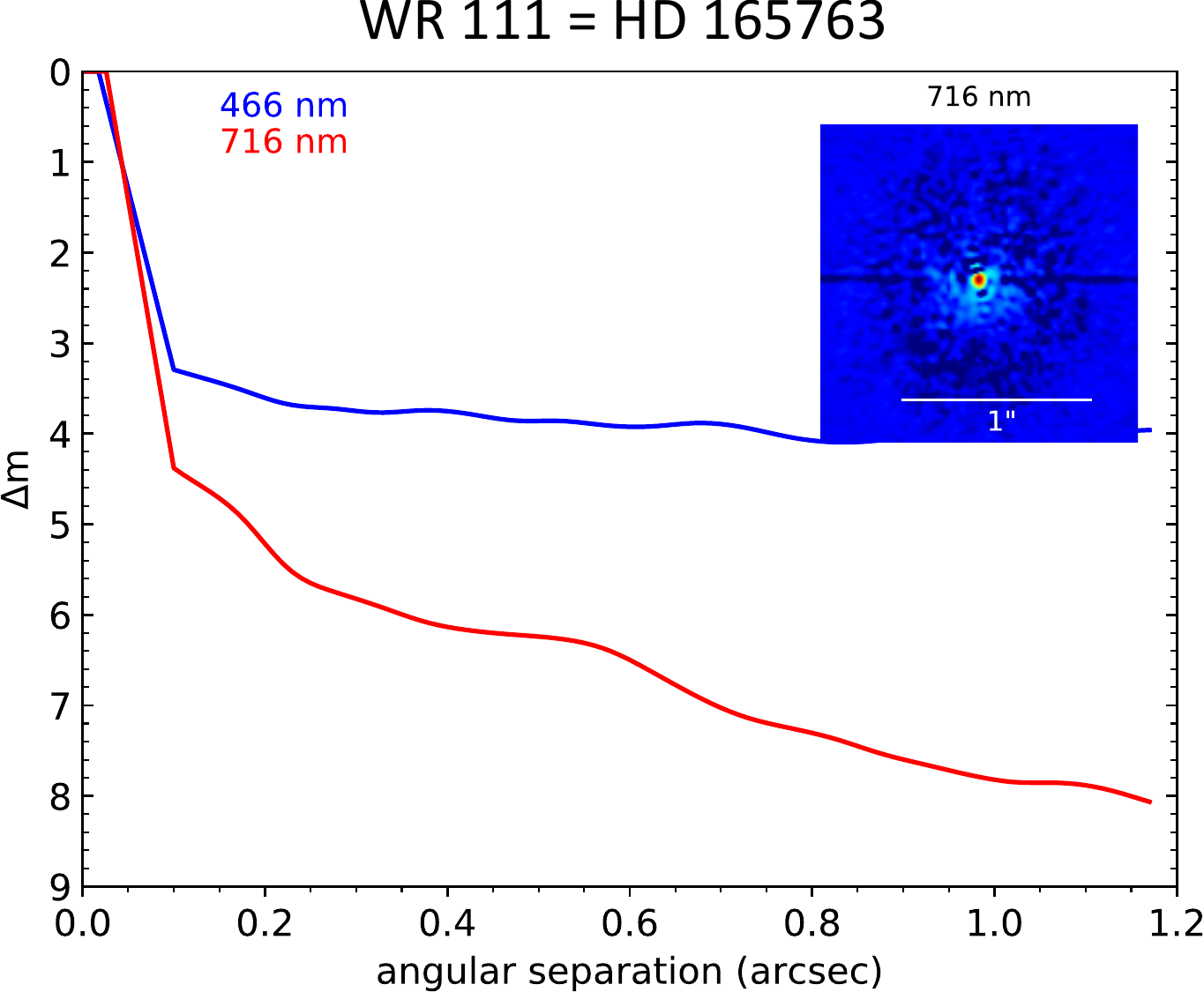}
\caption{{\bf Top:} The reconstructed speckle images and differential magnitude detection limits $\Delta$m of the Wolf-Rayet star WR 110 taken with the `Alopeke speckle camera at Gemini-N on 2020 Jun 11. There is no indication of any companion within 100 mas of WR 110 that is within $\sim$ 3 (5) magnitudes of the WR star in the 466 nm  (716 nm) narrowband filters. Neither is any companion detected within 4 (7) magnitudes in the 466 nm (716 nm) filters out to 1.2" from WR 110. 
{\bf Bottom:} The reconstructed speckle images and differential magnitude detection limits $\Delta$m of the WC5 Wolf-Rayet star WR 111 taken with the `Alopeke speckle camera at Gemini-N on 2020 June 11. There is no indication of any companion within 100 mas of WR 111 that is within $\sim$ 3 (4.5) magnitudes of the WR star in the 466 nm  (716 nm) narrowband filters. Neither is any companion detected within 4 (8) magnitudes in the 466 nm (716 nm) filters out to 1.2" from WR 111. See text for details.}\label{spectra}

\end{figure}
\clearpage

\clearpage
\begin{figure}

\hspace{4 cm}
\includegraphics[width=10 cm]{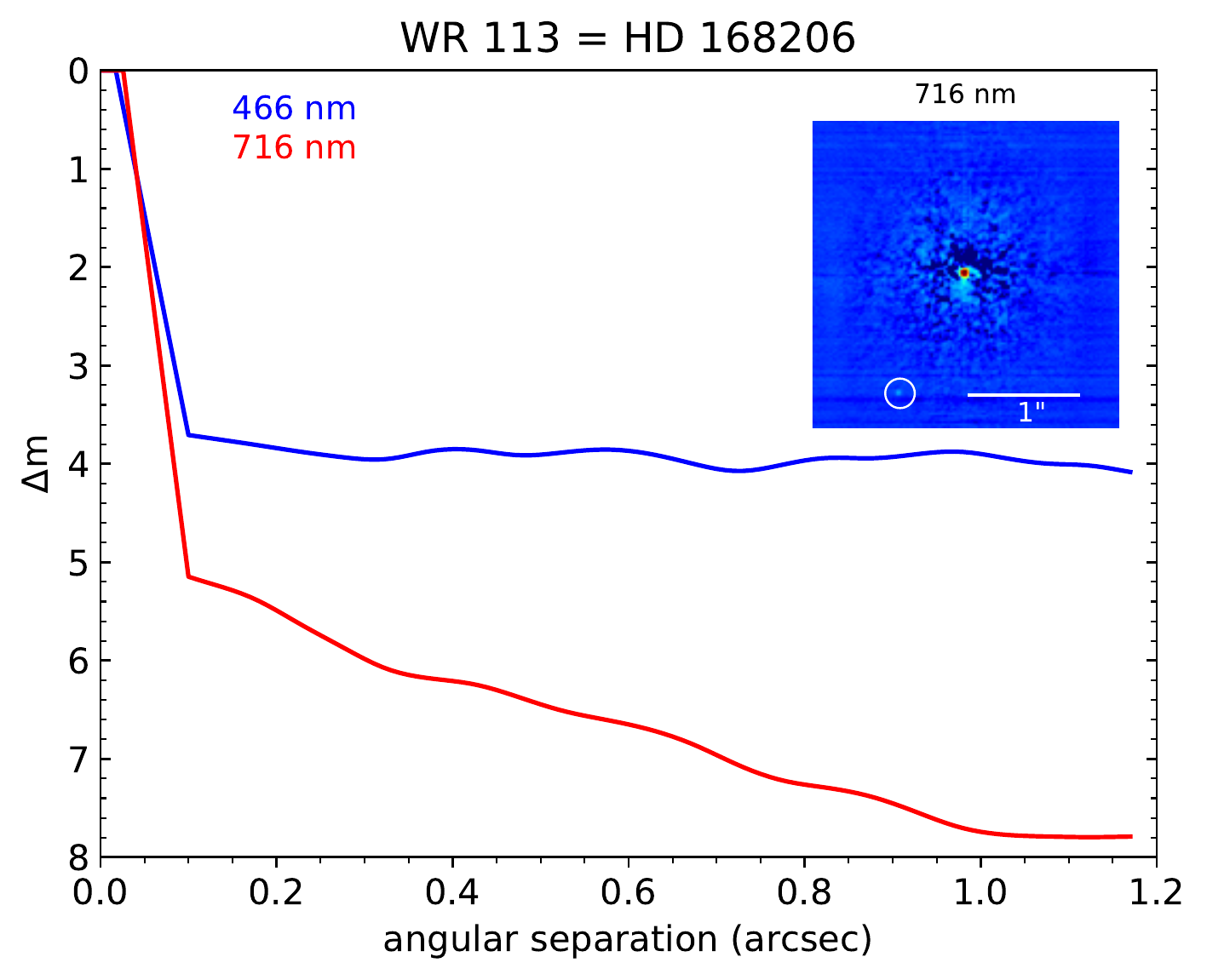}
\vspace{2mm}

\hspace{4 cm}
\includegraphics[width=10 cm]{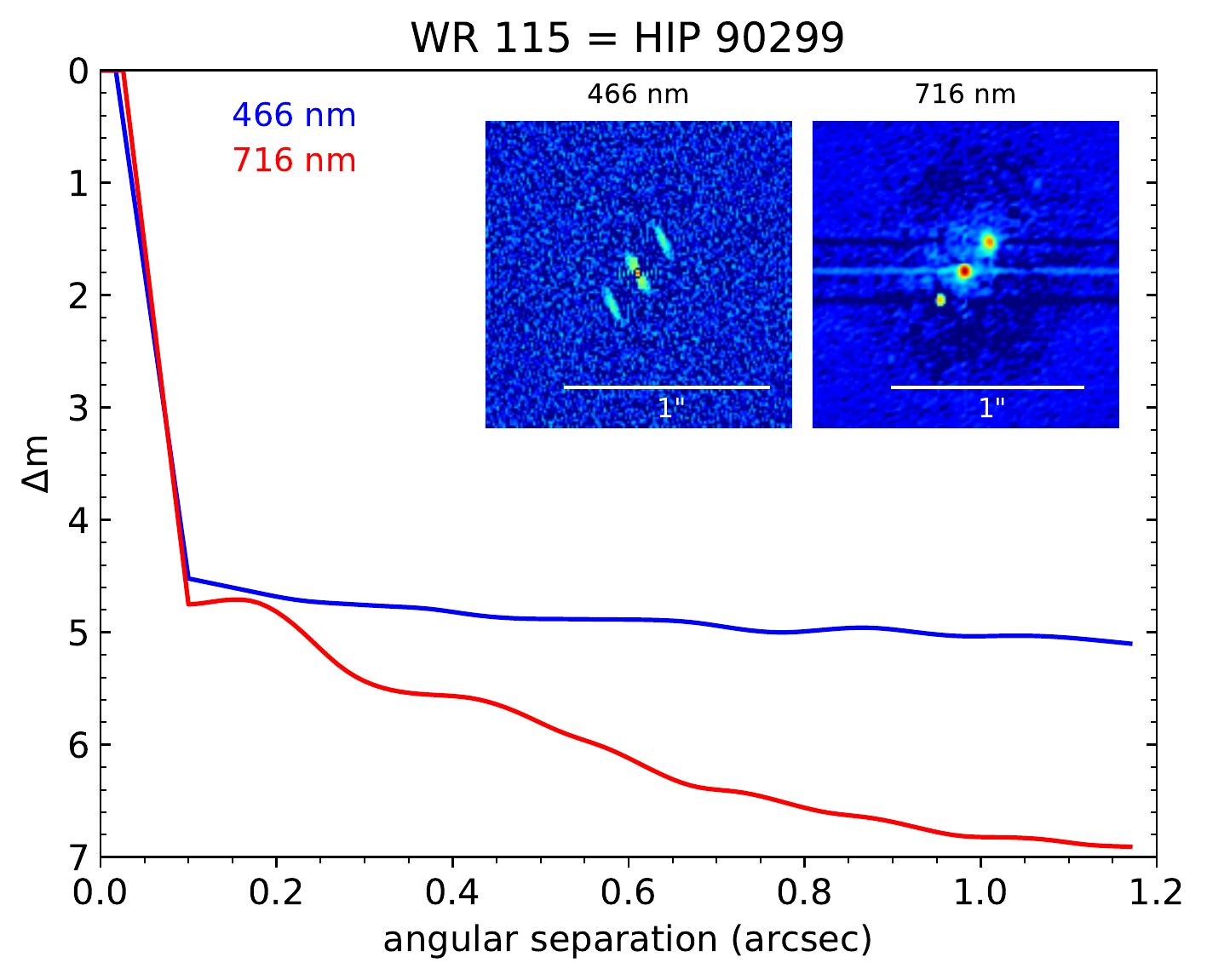}
\caption{{\bf Top:} The reconstructed speckle images and differential magnitude detection limits $\Delta$m of the WC8d + O8-9IV Wolf-Rayet star WR 113 taken with the `Alopeke speckle camera at Gemini-N on 2020 June 11. A star, circled in the figure and seen only in the 716 nm images, that is 8 magnitudes fainter than WR 113, is at an angular distance of 1.16" and position angle of 152 degrees measured from North through East.  The projected linear separation between the WR star and its companion is $\sim$ 2200 AU. {\bf Bottom:} The reconstructed speckle images and differential magnitude detection limits $\Delta$m of the WN6o Wolf-Rayet star WR 115 taken with the `Alopeke speckle camera at Gemini-N on 2020 June 11. A star that is 1.03 (1.60) magnitudes fainter than WR 115 is seen in the 716 (466) nm filter at an angular distance of 0.20" and a position angle of 321 degrees from the WR star. The projected linear separation between the WR star and its companion is $\sim$ 270 AU. The mirror image of that companion, seen below and to the left of the WR star is an artifact of the data reductions. See text for details.}\label{spectra}

\end{figure}

\clearpage

\clearpage
\begin{figure}

\hspace{4cm}
\includegraphics[width=10 cm]{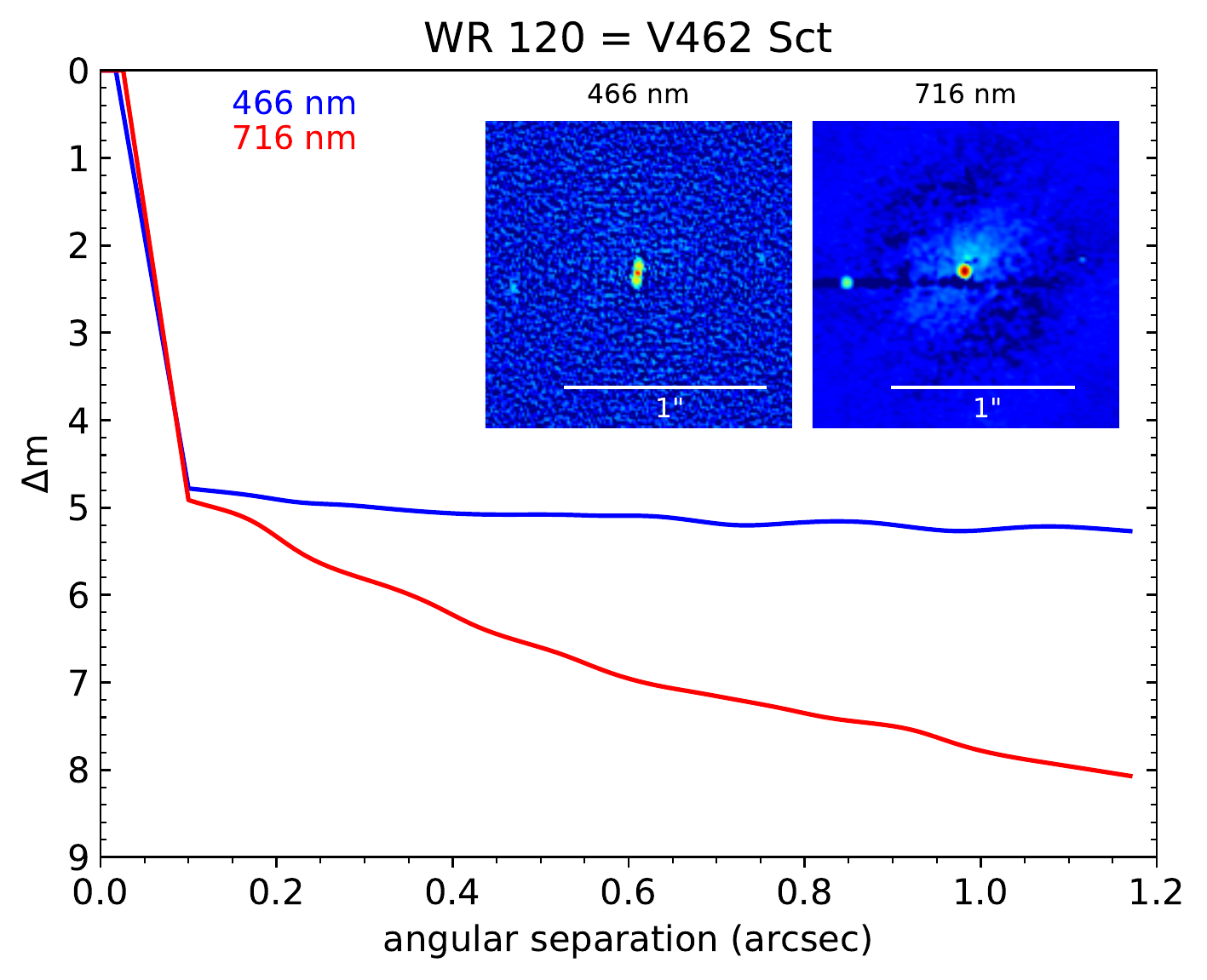}

\vspace{2 mm}
\hspace{4 cm}
\includegraphics[width=10 cm]{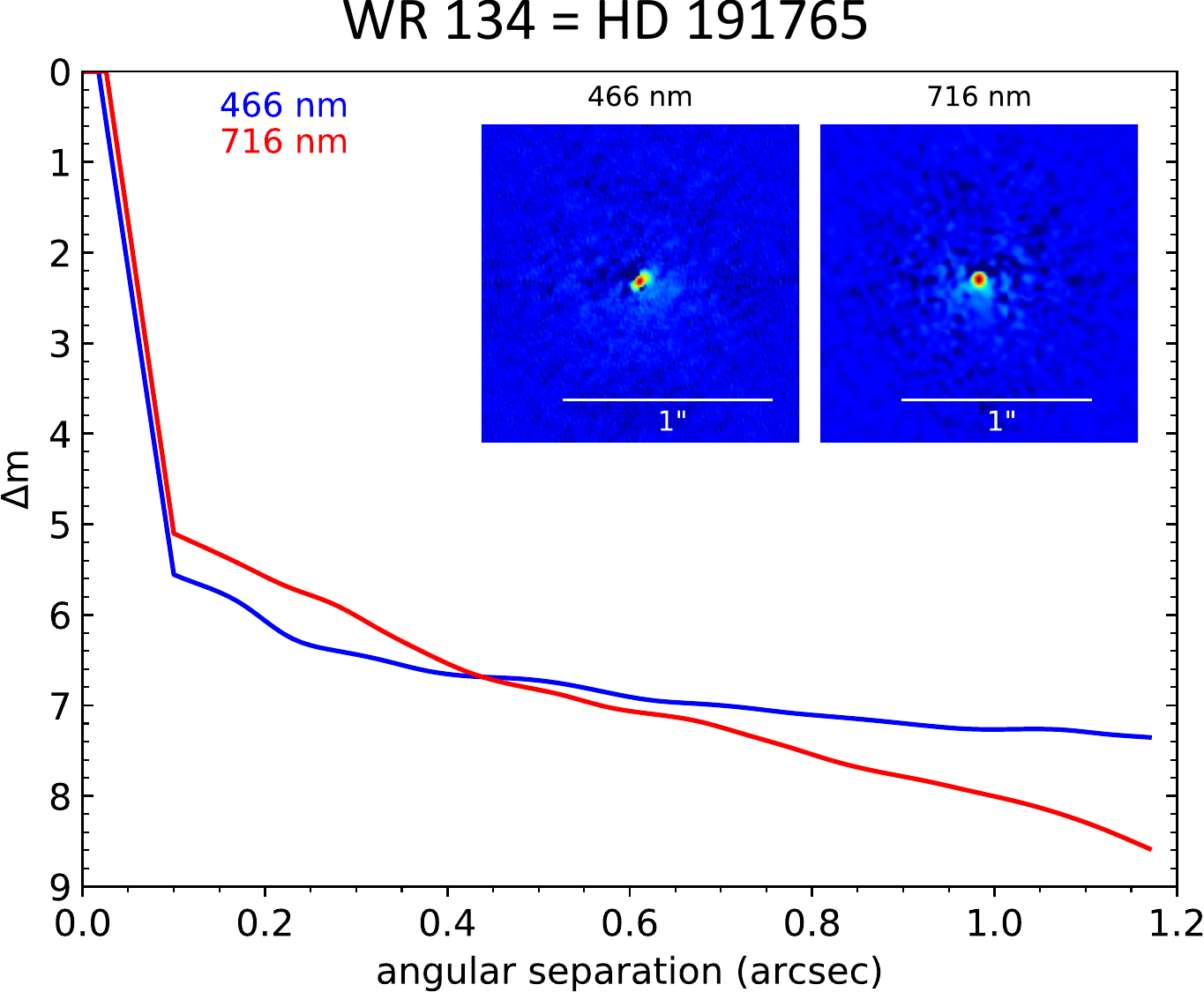}
\vspace{2mm}

\caption{{\bf Top:} The reconstructed speckle images and differential magnitude detection limits $\Delta$m of the WN7o Wolf-Rayet star WR 120 taken with the `Alopeke speckle camera at Gemini-N on 12 June 2020.  A star that is 2.11 (3.53) magnitudes fainter than WR 115 is seen in the 466 nm (716 nm) filter at an angular distance of 0.62" (0.66") and a position angle of 90 (96) degrees from the WR star. This corresponds to a projected linear separation of $\sim$ 1800 AU in the higher signal-to-noise red images. {\bf Bottom:} The reconstructed speckle images and differential magnitude detection limits $\Delta$m of the WN6b Wolf-Rayet star WR 134 taken with the `Alopeke speckle camera at Gemini-N on 2020 June 10. No companion is detected within 5 magnitudes of WR 134 at a separation $<$ 0.1" from the WR star; and no companion is detected within 7-8.4 mag of WR 134 in the separation range 0.1" $< \rho <$ 1.2". See text for details.}\label{spectra}

\end{figure}

\clearpage

\clearpage
\begin{figure}
\vspace{-2mm}
\hspace{4cm}
\includegraphics[width=10 cm]{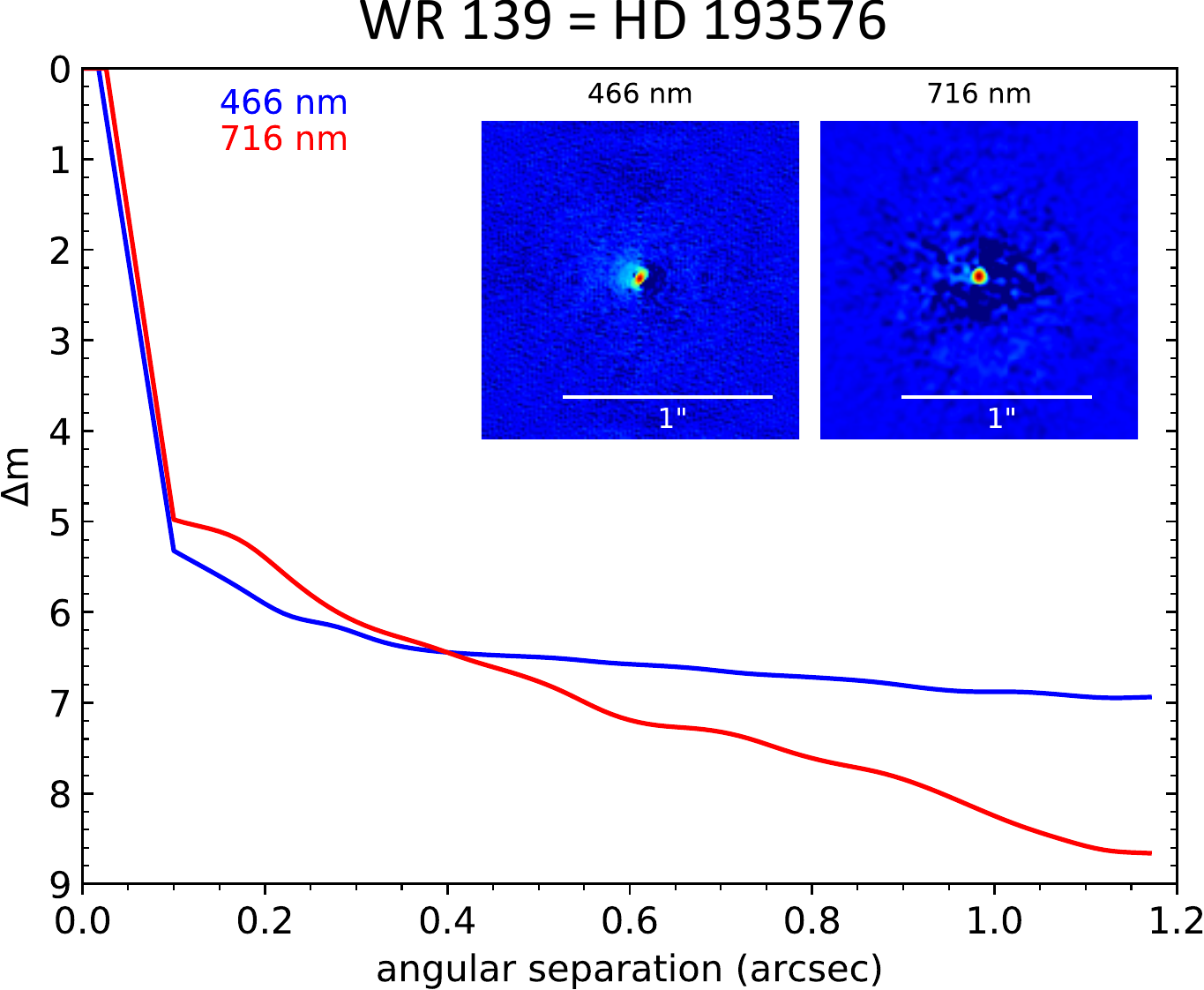}
\vspace{-1mm}

\hspace{4cm}
\includegraphics[width=10 cm]{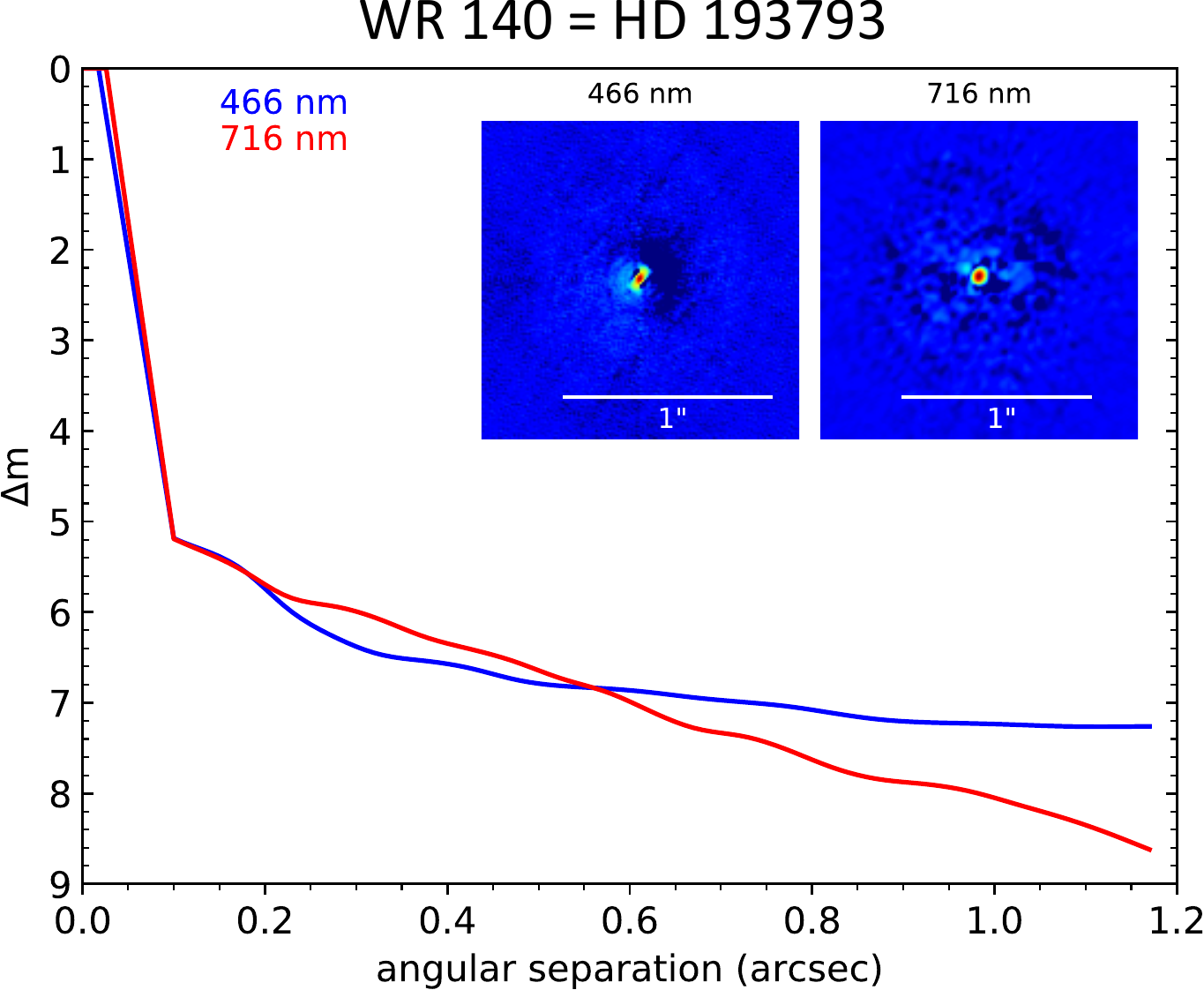}
\vspace{-5 mm}
\caption{{\bf Top:} The reconstructed speckle images and differential magnitude detection limits $\Delta$m of the WN5o + O6III-V Wolf-Rayet star WR 139 taken with the `Alopeke speckle camera at Gemini-N on 2020 June 12.  There is no indication of any companion within 100 mas of WR 139 that is within $\sim$ 5 magnitudes of the WR star in the 466 nm or 716 nm narrowband filters. Neither is any companion detected within 7 (8.6) magnitudes in the 466 nm (716 nm) filters out to 1.2 arcsec from WR 139. These angular separations corresponds to projected linear separations of  and  AU, respectively. {\bf Bottom:} The reconstructed speckle images and differential magnitude detection limits $\Delta$m of the WC7pd + O4-5 Wolf-Rayet star WR 140 taken with the `Alopeke speckle camera at Gemini-N on 2020 June 12.  The WR star and its companion were slightly closer than the 0.015" resolution limit of `Alopeke at the time of our observations. There is no indication of any companion within 100 milliarcsec of WR 140 that is within $\sim$ 5 magnitudes of the WR star in the 466 nm  or 716 nm narrowband filters. Neither is any companion detected within 7.2 (8.6) magnitudes in the 466 nm (716 nm) filters out to 1.2" from WR 140. These angular separations correspond to projected linear separations of  and  AU, respectively. See text for details.}\label{spectra}

\end{figure}

\clearpage

\begin{table*}
 \centering
  \caption{Gemini-N Speckle Targets: Eight WR stars}\label{Tid}
  \begin{tabular}{@{}rllllclc@{}}
  \hline
Star Name&Spec. type&b/v mag&Distance (kpc) \, & Proper Motion (mas/yr) &Observation UT Date \,&Integ. time (min) &\\
 
\hline

WR110=HD165688 &WN5-6b & 11.05/10.3 & 1.78 $\pm$ .07&1.90 $\pm$ 0.12 &2020-Jun-11 & 6\\
WR111=HD165763  &WC5 & 8.21/8.23 & 1.31 $\pm$ .08 &1.55 $\pm$ 0.17 &2020-Jun-11 &5 & \\
WR113=HD168206 & WC8d+O8-9IV & 9.89/9.43 & 1.90 $\pm$ .06&1.67 $\pm$ 0.10  & 2020-Jun-11 &5 & \\
WR115=HIP90299 & WN6o & 13.42/12.32 & $1.35_{-0.49}^{+1.13}$ &9.07 $\pm$1.87 &2020-Jun-11 &8 \\
WR120=V462 Sct & WN7o & 13.32/12.3 & 2.72 $\pm$ .34 &3.25 $\pm$ 0.33 &2020-Jun-12 &20\\
WR134=HD191765 & WN6b & 8.43/8.23 & 1.82 $\pm$ .06 &9.78 $\pm$ 0.07 &2020-Jun-10 &10\\
WR139=HD193576 & WN5o+O6III-V & 8.48/8.1 & 1.34 $\pm$ .04 &4.35 $\pm$ 0.07 &2020-Jun-12 &9\\
WR140=HD193793 & WC7pd+O5 & 7.34/7.07 & 1.78 $\pm$ .08 &5.08 $\pm$ 0.08 &2020-Jun-12 &9&\\

\hline
\end{tabular}\\
Spectral types and line-free b and v magnitudes \citep{smb68} from P. Crowther, https://pacrowther.staff.shef.ac.uk/WRcat/\\
Proper motions from \citet{rat20}\\
{\bf Distances from Bayesian priors \citep{rat20}, ZP corrections \citep{lin21} and Gaia EDR3 parallaxes \citep{gai21} \\}
\end{table*}

\begin{table*}
 \centering
  \caption{Gemini-N/`Alopeke Observations and Detected Companions of Eight WR stars}\label{Tid}
  \begin{tabular}{@{}rllllclc@{}}
  \hline
Star Name &Companion &$\theta$(degrees E of N)&$\rho$ (") &$\rho$ (AU)& Filter (nm)\\
 &$\Delta$mag&&&\\
\hline

WR110=HD165688 &---&---&---&---&466\\
WR110=HD165688 &---&---&---&---&716 \\
WR111=HD165763  &---&---&---&---&466\\
WR111=HD165763  &---&---&---&---&716 &\\
WR113=HD168206 &---&---&---&---&466 \\
WR113=HD168206 &8.0 $\pm$0.3&152&1.16&$\sim$2200&716\\
WR115=HIP90299 &1.60$\pm$0.3&320&0.20 &$\sim$270&466 \\
WR115=HIP90299 &1.03$\pm$0.15&321&0.20&$\sim$270&716 \\
WR120=V462 Sct &2.11$\pm0.3$&90&0.62&$\sim$1700&466\\
WR120=V462 Sct &3.53$\pm$0.15&96&0.66&$\sim$1800 &716 \\
WR134=HD191765 &--- &---&---&---& 466\\
WR134=HD191765 &--- &---&---&---& 716\\
WR139=HD193576 &--- &---&---&---&466\\
WR139=HD193576 & ---&---&---&---&716\\
WR140=HD193793 &---&---&---&---&466 \\
WR140=HD193793 &---&---&---&---&716\\

\hline
\end{tabular}\\
$\theta$ is the position angle of the companion relative to the WR star; uncertainties are $\pm$ 1 degree.

$\rho$ is the projected separation between the companion and the WR star; uncertainties are $\pm$ 1.5 mas. 

\end{table*}

\section{Summary and Conclusions} 
We carried out visible-light, narrowband speckle observations of eight of the nearest/brightest known WR stars to search for nearby companions, and to demonstrate the sensitivity of the latest speckle cameras with 8 meter telescopes to find new WR companions. Three new companions, to WR 113, WR 115 and WR 120 were found in the angular separation range $\sim$ 0.2" - 1.2", corresponding to projected linear separations of $\sim$ 2200, $\sim$ 270 and $\sim$ 1800 AU, respectively. 

If the companion to WR 113 is physically associated with the WR star, it would be (by far) the lowest luminosity companion known to any WR star. The surface density of stars similar in brightness to the companion of WR 113 suggests that there is a $\leq$ 0.5\% probability that it is a line-of-sight coincidence. Empirical surface densities of stars similar to the newly found companions to WR 115 and WR 120 suggest that they too are physical companions. Repeating these speckle observations in a few years will reveal (or not) relative motions between the WR stars and their companions, establishing or disproving physical associations.

This small survey re-enforces the view that close companions to WR stars are common, and many still await discovery. More importantly, especially via the 8-magnitude fainter companion of WR 113, we show that modern speckle cameras mounted on 8 meter class telescopes can detect hitherto undetectably close and faint companions, whose existence is suggested both by observations \citep{moe17} and predicted by recent numerical simulations \citep{wal19,cou21}. Knowing how often such faint and low mass stars accompany WR stars (and their O star progenitors) will be a strong constraint on massive star formation theories, and N-body simulations of young star clusters. 
\clearpage
\section*{Acknowledgments}
The data presented in this paper are based on observations obtained at the international Gemini Observatory, a program of NSF's NOIRLab, which is managed by the Association of Universities for Research in Astronomy (AURA) under a cooperative agreement with the National Science Foundation on behalf of the Gemini Observatory partnership: the National Science Foundation (United States), National Research Council (Canada), Agencia Nacional de Investigación y Desarrollo (Chile), Ministerio de Ciencia, Tecnología e Innovación (Argentina), Ministério da Ciência, Tecnologia, Inovações e Comunicações (Brazil), and Korea Astronomy and Space Science Institute (Republic of Korea). This work was enabled by observations made from the Gemini North telescope, located within the Maunakea Science Reserve and adjacent to the summit of Maunakea. We are grateful for the privilege of observing the Universe from a place that is unique in both its astronomical quality and its cultural significance. Observations in the paper made use of the High-Resolution Imaging instrument `Alopeke, which was funded by the NASA Exoplanet Exploration Program, built at the NASA Ames Research Center by Steve B. Howell, Nic Scott, Elliott P. Horch, and Emmett Quigley, and mounted on the Gemini-North telescope of the international Gemini Observatory. AFJM is grateful to NSERC (Canada) for financial aid. We thank the referee, Paul Crowther, for a thorough and thoughtful review which improved the paper, and for the WR star distances in Table 1. MMS thanks Mordecai Mac Low for discussions about binary star formation. We thank the Canadian Gemini Time Allocation Committee for excellent feedback and support, and their allocation of telescope time. The observations were obtained under Gemini proposal GN-2020A-Q-110. 
\section*{Data Availability Statement}

The data underlying this article will be shared on reasonable request to the corresponding author.

\newpage

\label{lastpage}

\end{document}